\documentclass[final,5p]{elsarticle}
\usepackage{amssymb}
\usepackage{mathrsfs}
\usepackage{mathtools}
\usepackage{hyperref}
\usepackage{pifont}
\usepackage{fontawesome}
\usepackage{stmaryrd}
\usepackage{enumerate}
\usepackage{multirow}
\usepackage{pdflscape}
\usepackage{adjustbox}

\hypersetup{
    colorlinks=true,
    linkcolor=blue,
    urlcolor=cyan,
}
\usepackage{array}
\usepackage{graphicx}
\usepackage{gensymb}
\usepackage{amsthm}
\usepackage{centernot}
\usepackage[dvipsnames]{xcolor}
\usepackage{tikz}
\usepackage[linesnumbered,lined,boxed,commentsnumbered]{algorithm2e}
\usepackage{appendix}
\usetikzlibrary{arrows,shapes.geometric}
\usetikzlibrary{calc}
\usepackage{caption}
\usepackage{subcaption}

\journal{Journal of the Franklin Institute}
\biboptions{sort&compress}

\SetCommentSty{mycommfont}

\newcommand\R{\mathbb{R}}

\newcommand\N{\mathbb{N}}

\renewcommand\d{\mathop{}\!\mathrm{d}}

\newcommand\sgn{\mathrm{sgn}}

\newcommand\dom{\text{dom }}

\DeclareMathOperator*{\argmin}{arg\,min}

\newtheorem{proposition}{Proposition}
\newtheorem{theorem}{Theorem}
\newtheorem{lemma}{Lemma}
\newtheorem{corollary}{Corollary}
\newtheorem{assumption}{Assumption}
\theoremstyle{definition}
\newtheorem{definition}{Definition}

\newtheorem{problem}{Problem}

\theoremstyle{remark}
\newtheorem{remark}{Remark}

\begin{document}
	\begin{frontmatter}
		\title{Explicit Time-Optimal Speed Profiles for Planar Paths with Monotone Curvature}
	
		\author[1]{Daniel Selvaratnam\corref{cor1}}
		\cortext[cor1]{Corresponding author\corref{cor1}}
		\ead{selv@kth.se}
		\author[2]{Michael Cantoni}
		\author[2]{Chris Manzie}

		\address[1]{Division of Decision and Control Systems, School of Electrical Engineering and Computer Science, KTH Royal Institute of Technology, 100 44 Stockholm, Sweden}
		\address[2]{Department of Electrical and Electronic Engineering, The University of Melbourne, Parkville, VIC 3010, Australia}
	
		\begin{abstract}
		Minimum-time speed profiles are constructed for planar paths with smooth strictly-monotonic signed curvature, subject to constraints on velocity, normal acceleration and tangential acceleration. The construction is explicit and exact, and global optimality is rigorously established from first principles under mild regularity conditions on the path. Free, fixed, and inequality-constrained boundary speeds are all accommodated. Numerical implementation is straightforward.
		\end{abstract}
	
	\end{frontmatter}
\section{Introduction}
Minimum-time traversal of a predefined path is a well-studied problem. In this paper, time-optimal speed profile synthesis is considered, without assuming a particular model for the agent \emph{kinetics}. Only \emph{kinematic} constraints on velocities and accelerations are imposed: maximum values for tangential acceleration and deceleration, normal acceleration, and speed, along with boundary conditions on the initial and terminal speeds. The resulting speed profile can, for example, be provided to a velocity-tracking controller on board a mobile robot, or used to predict travel times for path-planning. The novelty of the proposed approach is that it yields an explicit solution that is exact and globally optimal. To achieve this, the scope is restricted to planar paths that have smooth strictly-monotonic 	signed curvature. Although derived independently, these results correspond to a special case of the minimum-time velocity planning solution in \cite{consoliniSolutionMinimumtimeSpeed2020}, which involves the numerical solution of two ODEs with discontinuous right-hand sides. The restriction in scope enables exact construction of the solution instead, along with a first-principles proof of global optimality that relies only on standard tools from real analysis. By contrast, \cite{consoliniSolutionMinimumtimeSpeed2020} is underpinned by the more esoteric machinery of lattice theory. Note also that the solution method proposed here accommodates boundary speed constraints, which are not considered in \cite{consoliniSolutionMinimumtimeSpeed2020}. 


Early work on time-optimal velocity planning considers the motion of robotic manipulators, subject to limits on forces and torques at the manipulator joints~\cite{bobrowOptimalControlRobotic1983, bobrowTimeOptimalControlRobotic1985,shinMinimumtimeControlRobotic1985, pfeifferConceptManipulatorTrajectory1987,slotineImprovingEfficiencyTimeoptimal1989}. Given models for the manipulator kinetics, the force and torque limits lead to more complex acceleration constraints than the constants imposed here; however, those works do not consider velocity constraints. They are extended in \cite{xuanphuMethodOrientingCurves1997,lamirauxPathsTrajectoriesMultibody1998,zlajpahTimeOptimalPath1996,phamGeneralFastRobust2014,shenCompleteTimeOptimalPathConstrained2018} to accommodate velocity constraints, resulting in optimisation problems that capture the one herein as a special case. Note, \cite{xuanphuMethodOrientingCurves1997} provides analysis based on the method of orienting curves, together with a suggested solution procedure that is not implemented. Given a path parameterised by a scalar $s$, computation of maximum velocity curves in the $s \dot{s}$-plane via numerical
search lies at the core of \cite{lamirauxPathsTrajectoriesMultibody1998,zlajpahTimeOptimalPath1996,phamGeneralFastRobust2014,shenCompleteTimeOptimalPathConstrained2018}. The proposed algorithms do not yield explicit solutions, and their implementation can present numerical difficulties \cite{phamGeneralFastRobust2014,phamNewApproachTimeOptimal2018}. Alternative numerical approaches rely on path discretisation~\cite{verscheureTimeOptimalPathTracking2009,lippMinimumtimeSpeedOptimisation2014,consoliniOptimalComplexityAlgorithm2017,phamNewApproachTimeOptimal2018}, or dynamic programming~\cite{shinDynamicProgrammingApproach1986,singhOptimalTrajectoryGeneration1987}.

The only prior works known to the authors that propose analytical solutions to the minimum-time velocity planning problem are \cite{consoliniSolutionMinimumtimeSpeed2020} and \cite{velenisOptimalVelocityProfile2005}. 
The latter presents an explicit time-optimal velocity profile for a point-mass vehicle subject to a net force constraint. A simple kinetic model for the normal and tangential force components is adopted, and an upper bound imposed on the magnitude of the vector sum. This leads to a mixed constraint, involving both velocity and tangential acceleration in the same equation. An optimal control problem is formulated, and Pontryagin's Maximum Principle (PMP) applied to derive a solution. Herein, by contrast, the normal and tangential acceleration constraints are decoupled. The corresponding optimal control problem involves pure state inequality constraints, which demand more advanced techniques than in the case of mixed constraints~\cite[Section 4]{peschPracticalGuideSolution1994}. Depending on the path, PMP may require a large number of jump conditions to be resolved, which hinders the discovery of an explicit solution. The approach of \cite{consoliniSolutionMinimumtimeSpeed2020} avoids this issue by invoking lattice theory to establish the optimal solution as the pointwise minimum of the solutions to two discontinuous ODEs. The first-principles approach developed here also avoids jump conditions, and though established independently, can be viewed as specialising \cite{consoliniSolutionMinimumtimeSpeed2020} to paths with strictly-monotone signed curvature under constant velocity and acceleration limits.

Paths with strictly-monotone signed curvature constitute an important class of curves. Clothoids are used for highway and railway track design~\cite{levienEulerSpiralMathematical2008}, and in path-planning for autonomous vehicles~\cite{limaClothoidBasedSpeedProfiler2015,chenAccurateEfficientApproximation2017,silvaClothoidBasedGlobalPath2018}, because they have linear signed curvature. This allows vehicles to transition smoothly between two paths. A method for generating feasible velocity profiles along clothoids is proposed in \cite{limaClothoidBasedSpeedProfiler2015}, but time-optimality is not claimed. Semi-analytical time-optimal velocity profiles are derived in \cite{fregoSemianalyticalMinimumTime2017} specifically for clothoids, based on PMP. 
Since the Cartesian parametric equations of a clothoid require the evaluation of Fresnel Integrals, they are often approximated with polynomial series expansions instead~\cite{levienEulerSpiralMathematical2008,chenAccurateEfficientApproximation2017}. The strict monotonicity assumption on path curvature here admits both clothoids and their polynomial approximations.  
Beyond these, any path of practical use can be decomposed into a finite number of segments with either strictly-monotonic or constant signed curvature. Since velocity profiling for the latter is trivial, and velocity profiling for the former is treated here, a time-optimal velocity profile for the entire path can be constructed by choosing the right boundary conditions for each segment. This reduces an infinite-dimensional optimisation problem to a finite-dimensional one, without introducing any approximations. Such a strategy may be more efficient than solving the discontinuous ODEs in \cite{consoliniSolutionMinimumtimeSpeed2020} numerically, so extensions in this direction are the subject of ongoing work. 

\section{Preliminaries}
\subsection{Notation}
The (non-strict) subset relation is denoted by $\subset$. The function $g\circ f:F \to H$ denotes the composition of $f:F \to G$ with $g:G \to H$, and
$f^{-1}[Y] \subset F$ the preimage of $Y \subset G$ under $f$. The real numbers are denoted by $\R$, and the natural numbers by $\N := \{0,1,...\}$. For $i,j \in \N$, $ [i:j]:= \{k \in \N \mid i \leq k \leq j \}$. By convention, $\inf \emptyset := \infty$ and $\sup \emptyset:= - \infty$. For $n \in \{1,2,\hdots\}$, elements of $\R^n$ are considered column vectors. Given $a<b$, the derivative of $h:[a,b] \to \R^n$ at $s \in [a,b]$ is denoted $h'(s)$ wherever it exists. The right derivative of $h$ at $s$ is given by
$$ h^+(s):= \lim_{\tau \downarrow 0} \frac{h(s+\tau) - h(s)}{\tau},$$
wherever the one-sided limit exists. Observe, $h'(a) = h^+(a)$ at the left endpoint,.
\subsection{Technical results}
Subsequent analysis hinges on the following propositions about continuous functions. Proofs are provided for completeness.
\begin{proposition} \label{prop:deriv}
	For any continuously differentiable $f:[0,1) \to \R$,
	$$ \limsup_{s \to 1} f(s) = \infty \implies \limsup_{s \to 1} f'(s) = \infty.$$
\end{proposition}
\begin{proof}
	If $\limsup_{s \to 1} f'(s)< \infty$, then $c:= \sup\{ f'(s) \mid s \in [0,1)\} < \infty$, and
	\begin{align*}
		\forall s \in [0,1),\ f(s) &= f(0) + \int_0^s f'(\tau) \d\tau \\
		& \leq f(0)  + \int_0^s c \d\tau = f(0) + cs,
	\end{align*}
	by which$$\limsup_{s \to 1} f(s) \leq \lim_{s \to 1} \Big( f(0) + cs\Big) = f(0) + c < \infty.$$
\end{proof}
\begin{proposition} \label{prop:inf}
	Let $f:[0,1) \to \R$ be continuous, and $$ a:= \inf \{ s \in [0,1) \mid f(s) > 0\}.$$ If $a< \infty$, then $a\in [0,1)$ and $f(a) \geq 0$. Moreover, if $0 < a< \infty$, then $f(s) \leq 0$ for all $s \in [0,a]$, and in particular, $f(a) = 0$.
\end{proposition}
\begin{proof}
	Let $S:= \{ s \in [0,1) \mid f(s) > 0\} = f^{-1}[(0,\infty)] $. Since $S \subset [0,1)$, $a= \inf S \geq \inf [0,1) = 0$. If $S = \emptyset$, then $a= \infty$, by convention. Suppose that $a< \infty$, which implies $S \neq \emptyset$. Then there exists $\tau \in [0,1)$ such that $f(\tau) > 0$. It follows that $a\leq \tau < 1$, and thus, $a\in [0,1) = \dom f$.

	Now, let $T:= \{ s \in [0,1) \mid f(s) < 0\} = f^{-1}[(-\infty,0)]$. Both $S$ and $T$ are open relative to $[0,1)$, because $f$ is continuous. Suppose $a\in T$. Then there exists a neighbourhood $U \subset [0,1)$ of $a$ such that $U \subset T$. Since $S$ and $T$ are disjoint, $U \cap S = \emptyset$. But $a= \inf S$ is a boundary point of $S$ by definition, which implies the contradiction $U \cap S \neq \emptyset$. Thus, $a\notin T$.

	Finally, to again obtain a contradiction, suppose that $f(\tau) > 0$ for some $\tau \in [0,s] \subset [0,1)$, where $s>0$. Then $\tau \in S$ by definition, which implies $a= \inf S \leq \tau$. Since $\tau \leq s$ by hypothesis, $ a= \tau \in S$, and thus, $\inf S \in S$. But $S$ is open and therefore contains none of its boundary points, including $\inf S$, leading to a contradiction.
\end{proof}
\begin{remark}
	The preceding results extend to any $f:[a,b) \to \infty$, with $a<b$, because $[a,b)$ can be mapped smoothly to $[0,1)$; e.g., $s \mapsto \frac{s-a}{b-a}$.
\end{remark}
\begin{proposition} \label{prop:derivAC}
	Let $a<b$, and $f,g:[a,b] \to \R$ be Lipschitz. Then $h:[a,b] \to \R,\ h(s):= \min\{f(s),g(s)\}$ is also Lipschitz. Let $F \subset \R$ be such that $f'(s) \in F$ for almost every $s \in [a,b]$, and $G$ such that $g'(s) \in G$ for almost every $s \in [a,b]$. If $f$ is strictly increasing and $g$ non-increasing, then
	\begin{equation} h'(s) \in F \cup G \label{eq:h'} \end{equation}
	for almost every $s \in [a,b]$. The same holds if $f$ is strictly decreasing and $g$ non-decreasing.
\end{proposition}
\begin{proof}
	That $h$ is Lipschitz follows from \cite[Proposition 2.3.9]{cobzasLipschitzFunctions2019}. If $f$ is strictly increasing and $g$ non-increasing, then there is at most one point at which $f(s) = g(s)$. The same holds if $f$ is strictly decreasing and $g$ non-decreasing. In either case, the set $E:= \{ s \in [a,b] \mid f(s) = g(s) \}$ is a singleton.

	Let $S^c$ denote $[a,b] \setminus S$ for any $S \subset [a,b]$. Define $\hat{F}:= \{ s \in [a,b] \mid f'(s) \in F\}$ and $\hat{G}:= \{ s \in [a,b] \mid g'(s) \in G\}$. Choose any $s \in \hat{F} \cap \hat{G} \cap E^c $, recalling that $f$ and $g$ are both continuous. Since $f(s) \neq g(s)$, either $f(s) < g(s)$ or $g(s) < f(s)$. If the former, then there exists a neighbourhood $U \subset [a,b]$ of $s$ such that $h(\tau) = f(\tau)$ for all $\tau \in U$, and therefore $h'(s) = f'(s) \in F$. If the latter, then $h'(s) = g'(s) \in G$ for the same reasons. This establishes \eqref{eq:h'} for all $s \in \hat{F} \cap \hat{G} \cap E^c$. If \eqref{eq:h'} does not hold, then $ s\in (\hat{F} \cap \hat{G} \cap E^c)^c = \hat{F}^c \cup \hat{G}^c \cup E$, which has measure zero by hypothesis.
\end{proof}
\section{Problem formulation}
Consider a thrice continuously differentiable curve $r:[0,L] \to \R^2,$ parametrised by arc-length $s$.
Its signed curvature $$\kappa:[0,L] \to \R,\ \kappa(s)=  \det \begin{bmatrix} r'(s) & r''(s) \end{bmatrix}$$
is then continuously differentiable, and $\|r'(s)\|=1$ for all $s \in [0,L]$. The goal of this paper is to construct a continuous speed profile $v:[0,L] \to \R$ that minimises travel-time along $r$, subject to a finite maximum tangential acceleration $A>0$, maximum braking deceleration $B>0$, maximum normal acceleration $C>0$, and maximum speed $V>0$.
This is encoded in the optimisation problem below, where $v_0,v_L \geq 0$ are initial and terminal speeds, and $x(s)= v(s)^2$ is the squared-speed at distance $s \in [0,L]$ along the path. The tangential acceleration at $s$ is then given by
$ v(s)v'(s) = \frac{1}{2} \frac{ \d (v(s)^2)}{\d s} =  \frac{x'(s)}{2}$, and the normal acceleration by $ x(s)|\kappa(s)|$. The optimisation problem is cast in terms of $x$ rather than $v$, as is standard in the literature~\cite{verscheureTimeOptimalPathTracking2009,lippMinimumtimeSpeedOptimisation2014,consoliniSolutionMinimumtimeSpeed2020}.
\begin{problem}[Time-optimal speed-squared profile]\label{prob:OCP}
Given $A,B,C,V,L> 0$, $\kappa:[0,L] \to \R$ continuously differentiable, and $v_0,v_L \geq 0$, find an absolutely continuous $x:[0,L] \to [0,\infty)$ that minimises
	\begin{equation} J(x) = \int_0^L \dfrac{1}{ \sqrt{ x(s)}} \d s, \label{eq:cost} \end{equation}
	subject to
		\begin{equation}
		-2B \leq x'(s) \leq 2A \label{eq:ocpDynamics}
	\end{equation}
for almost every $s \in [0,L]$, and
		\begin{align}
			&x(0) \leq v_0^2 \label{eq:initBC} \\
			&x(L) \leq v_L^2 \label{eq:finalBC}\\
		\forall s \in [0,L],\ &x(s)  \leq \min \left\{ \frac{C}{|\kappa(s)|},V^2 \right \}. \label{eq:stateConstraints}
	\end{align}
\end{problem}
\begin{definition}[Feasibility] \label{def:feasible}
	An absolutely continuous $x:[0,L] \to [0,\infty)$ is \emph{feasible} iff \eqref{eq:ocpDynamics} holds for almost every $s \in [0,L]$, and it satisfies \eqref{eq:initBC}--\eqref{eq:stateConstraints}.
\end{definition}
\begin{remark}[Boundary conditions] \label{rem:BCs}
Inequality constraints are imposed on the boundary values in Problem \ref{prob:OCP}, because unconstrained and equality-constrained boundary values can also be addressed within this framework. To make the initial speed free, set $v_0 \geq \min\{\sqrt{C|\kappa(0)|^{-1}},V\}$, and setting $v_L \geq \min\{\sqrt{C|\kappa(L)|^{-1}},V\}$ makes the final speed free. It is ultimately established in Corollary \ref{cor:BCs} that the solution proposed in Theorem \ref{thm:soln} satisfies \eqref{eq:initBC} and \eqref{eq:finalBC} with equality, if such a feasible profile exists.
\end{remark}
Note that absolutely continuous functions are differentiable almost everywhere, and satisfy the Fundamental Theorem of Calculus~\cite[Chapter 6.5]{roydenRealAnalysis2010}. An explicit solution to Problem \ref{prob:OCP} is constructed next, under the following assumption.
\begin{assumption} \label{ass:curvature}
	The signed curvature $\kappa:[0,L] \to \R$ is strictly-monotonic.
\end{assumption}
\section{Construction of an optimal speed-squared profile}
\begin{algorithm*}[t]
	\SetAlgoLined
	\DontPrintSemicolon
	\SetKwFunction{Fxplus}{Sweep}
	\SetKw{fset}{set}
	\SetKwProg{Fn}{Function}{:}{end}
	\Fn(){\Fxplus{$\kappa,A,C,v_0$}}{
		\KwIn{$\kappa:[0,L] \to \R$, $A > 0$, $C>0$, $v_0 \geq 0$}
		\KwOut{$x_F:[0,L] \to [0,\infty]$}
		\BlankLine
		$\displaystyle s_0 :=\argmin_{s \in [0,L] }|\kappa(s)|$ \label{ln:t0} \tcp{Minimiser is unique under Assumption \ref{ass:curvature}}
		$c_0 := 0$, $k:=0$ \label{ln:initWhile}\;
		\If(\texttt{\textcolor{blue}{\footnotesize // If initial speed below curvature boundary}}){$v_0^2 < C|\kappa(0)|^{-1}$}{
			$a_0 := 0$ \label{ln:a0} \tcp{Start at max acceleration}
			\fset $x_0:[0,L] \to \R,\ x_0(s):= v_0^2 + 2A s $ \label{ln:x0}\tcp{Initial max acceleration profile}
			$c_1 := \inf \left \{ s \in [0,s_0) \mid  C|\kappa(s)|^{-1} < x_0(s) \right\}$ \label{ln:c1} \tcp{First switch to curvature constraint}
			$k:=1$
		}
		\While{$c_k<\infty$}{\label{ln:startWhile}
			$a_k := \inf \left \{ s \in [c_k,s_0) \mid \frac{C |\kappa'(s)|}{\kappa(s)^2} > 2A \right\}$ \label{ln:ak} \tcp{Switch to max acceleration}
			\eIf{$a_k<\infty$}{
				\fset $x_k:[0,L] \to \R,\ x_k(s) :=  C |\kappa(a_k)|^{-1} + 2A\cdot(s - a_k)$ \label{ln:x+k}	\tcp{Max acceleration profile}
				$c_{k+1} := \inf \left \{ s \in [a_k,s_0) \mid  C|\kappa(s)|^{-1} < x_k(s) \right\}$ \label{ln:ck} \tcp{Switch to curvature constraint}
			}{$c_{k+1} := a_k$}
			$k \leftarrow k+1$
		} \label{ln:endWhile}
		\fset $x_F:[0,L] \to [0,\infty],\ x_F(s):= \begin{cases}
			C |\kappa(s)|^{-1},& \text{if } s \in \bigcup_{n=0}^{k-1} [c_n,a_n) \\
			x_{\max\{ n \mid c_n \leq s\}}(s),& \text{otherwise}
		\end{cases}$ \label{ln:x+}\;
		\KwRet $x_F$ }
	\caption{Constructs a speed-squared profile $x_F$ given an initial speed $v_0$, maximum tangential and normal acceleration limits $A$ and $C$, respectively, and path curvature $\kappa$ parametrised by arc-length.}
	\label{alg:xplus}
\end{algorithm*}
Algorithm \ref{alg:xplus} is used to explicitly construct a solution to Problem \ref{prob:OCP}. It is invoked twice to process curvature information in two `sweeps' along the path. The forward sweep progresses in the direction of increasing $s$ to yield a speed-squared profile $x_F:[0,L] \to [0,\infty]$ that switches between the curvature constraint boundary associated with \eqref{eq:stateConstraints}, and the maximum tangential \emph{acceleration} boundary in \eqref{eq:ocpDynamics}. Line \ref{ln:ak} defines the points $a_k$ at which it enters the latter, and Line \ref{ln:ck} the points $c_k$ at which it switches back to the former. Pathological paths requiring an infinite number of switches over a finite distance are excluded by the following assumption. (Cf. \cite[Hypothesis 2]{bobrowTimeOptimalControlRobotic1985}, the final paragraph of \cite[Section 2]{xuanphuMethodOrientingCurves1997}, and the discussion immediately preceding \cite[Theorem 1]{shinMinimumtimeControlRobotic1985}.)
\begin{assumption}[Forward sweep termination] \label{ass:finiteA}
The given constants $A,C > 0$ and curvature $\kappa$ in Problem \ref{prob:OCP} are such that $\mathtt{Sweep}(\kappa,A,C,v_0)$ returns a result; i.e., the while-loop (Lines \ref{ln:startWhile}--\ref{ln:endWhile}) in Algorithm \ref{alg:xplus} terminates.
\end{assumption}
The reverse sweep progresses in the opposite direction to yield a profile $x_R:[0,L] \to [0,\infty]$ that switches between the curvature constraint boundary and maximum tangential \emph{deceleration} boundary. A similar path curvature regularity assumption is needed to guarantee termination of the algorithm.
\begin{assumption}[Reverse sweep termination] \label{ass:finiteB}
	Define the transformation $\gamma:[0,L] \to [0,L],\ \gamma(s) := L -s$. The given constants $B,C > 0$ and curvature $\kappa$ in Problem \ref{prob:OCP} are such that $\mathtt{Sweep}(\kappa\circ \gamma,B,C,v_L)$ returns a result; i.e., the while-loop (Lines \ref{ln:startWhile}--\ref{ln:endWhile}) in Algorithm \ref{alg:xplus} terminates.
\end{assumption}
The pointwise minimum of $x_F$, $x_R$, and $V^2$, is an optimal profile. This result is the main contribution of the paper.
\begin{theorem} \label{thm:soln}
Under Assumptions \ref{ass:curvature}--\ref{ass:finiteB}, define
\begin{align*}
	x_F&:=\mathtt{Sweep}(\kappa,A,C,v_0), \\
	x_R &:= \mathtt{Sweep}(\kappa \circ \gamma,B,C,v_L)\circ \gamma.
\end{align*}
 The speed-squared profile
	$$ x^\star:[0,L] \to [0,\infty),\ x^\star(s):= \min\{ x_F(s),x_R(s),V^2\}.$$
	solves Problem \ref{prob:OCP}.
\end{theorem}
\begin{remark}\label{rem:lipschitz}
	Lemma \ref{lem:feasible} establishes that the proposed solution $x^\star$ is Lipschitz, which implies (but is not necessary for) the absolute continuity required in Definition \ref{def:feasible}. This stronger property is desirable, because it ensures the tangential acceleration is bounded wherever it exists. It comes without sacrificing optimality.
\end{remark}
The remainder of this section is devoted to proving Theorem \ref{thm:soln}. This involves verifying that the profiles $x_F$ and $x_R$ generated by Algorithm \ref{alg:xplus} are well defined, before establishing the feasibility and global optimality of $x^\star$.
\subsection{The point of minimum unsigned curvature}
Assumption \ref{ass:curvature} requires the signed curvature $\kappa$ to be either strictly increasing or strictly decreasing. The unsigned curvature $|\kappa(s)|$ may, however, have a turning point. Line \ref{ln:t0} of Algorithm \ref{alg:xplus} defines $s_0 \in [0,L]$ as the minimiser of unsigned curvature along the path:
\begin{equation} s_0:=\argmin_{s \in [0,L] }|\kappa(s)|. \label{eq:t0} \end{equation}
The forward sweep deals with the decreasing portion of $|\kappa(s)|$ over $s \in [0,s_0]$, and the reverse sweep with the increasing portion over $s \in [s_0,L]$.
\begin{lemma} \label{lem:uniqueMin}
Under Assumption \ref{ass:curvature}, there exists a unique minimiser of $|\kappa(s)|$ over $s \in [0,L]$.
\end{lemma}
\begin{proof}
The continuity of $\kappa$ and the compactness of $[0,L]$ guarantee the existence of a minimiser, by the extreme value theorem of Bolzano and Weiestrass. Suppose $\kappa$ changes sign. Then there exists $s_0 \in (0,L)$ such that $\kappa(s_0) = 0$ by the intermediate value theorem. It is unique by Assumption \ref{ass:curvature}. Thus, $s_0$ is the unique minimiser of $|\kappa(s)|$. Suppose now that $\kappa$ does not change sign, and let $s_0,s_1 \in [0,L]$ be minimisers of $|\kappa(s)|$. Then $\lvert \kappa(s_0) \rvert = \lvert \kappa(s_1) \rvert$, and since $\kappa(s_0)$ and $\kappa(s_1)$ have the same sign, it follows that $\kappa(s_0) = \kappa(s_1)$. If $s_0 \neq s_1$, this contradicts the strict monotonicity of $\kappa$.
\end{proof}
The value $\kappa(s_0)$ can be non-zero only if $s_0$ is at the start or end of the path, in which case the unsigned curvature $|\kappa(s)|$ is also strictly-monotonic, as demonstrated below.
\begin{lemma} \label{lem:kappa=0}
	Under Assumption \ref{ass:curvature}, if $s_0\in (0,L)$, then $\kappa(s_0) = 0$.
\end{lemma}
\begin{proof}
Suppose $s_0 \in (0,L)$ and $\lvert \kappa(s_0) \rvert > 0$. Since $\kappa$ is continuous, there exists a neighbourhood $U \subset (0,L)$ of $s_0$ such that $\lvert \kappa(s) \rvert > 0$ for all $s \in U$. The strict monotonicity of $\kappa$ then implies the existence of some $s_1 \in U$ such that $\lvert \kappa(s_1) \rvert < \lvert \kappa(s_0) \rvert$, which contradicts the definition of $s_0$.
\end{proof}
\begin{corollary} \label{cor:kappaNonzero}
Under Assumption \ref{ass:curvature}, if $\kappa(s_0) \neq 0$, then $s_0 \in \{0,L\}$.
\end{corollary}
\begin{remark} \label{rem:endpoint0}
The map $s \mapsto |\kappa(s)|$ is strictly decreasing iff $s_0 = L$, and strictly increasing iff $s_0 = 0$.
\end{remark}

The maximum speed-squared permitted by the normal acceleration constraint at $s \in [0,L]$ is given by $\xi(s) := C|\kappa(s)|^{-1}$. Below is an expression for its derivative over the initial portion of the path $[0,s_0)$, which has decreasing unsigned curvature.
\begin{lemma} \label{lem:vdot}
Suppose $s_0 > 0$ and let $\xi:[0,s_0)  \to \R,\ \xi(s):= \frac{C}{|\kappa(s)|}$. Under Assumption \ref{ass:curvature},
$$ \forall s \in [0,s_0),\ \xi'(s) =  \frac{C |\kappa'(s)|}{\kappa(s)^2}.$$
\end{lemma}
\begin{proof}
Applying the chain rule, $\xi'(s) =  - \sgn(\kappa(s))\frac{C \kappa'(s)}{\kappa(s)^2}$
for all $s < s_0$. By Lemma \ref{lem:uniqueMin},
\begin{equation} \forall s < s_0,\ \lvert \kappa(s) \rvert > \lvert \kappa(s_0) \rvert \geq 0. \label{eq:t0_properties} \end{equation}
Suppose $\kappa$ is strictly increasing. Then $\kappa(s) < 0$ for all $s < s_0$, by which $- \sgn(\kappa(s)) = 1$.
Moreover, $\kappa'(s) \geq 0$, which implies $\kappa'(s) = |\kappa'(s)|$.
If, instead, $\kappa$ is strictly decreasing, then $\kappa'(s) = -\lvert\kappa'(s) \rvert \leq 0$, and $\kappa(s) >0$ for all $s < s_0$, by which $- \sgn(\kappa(s)) = -1$. In both cases, $- \sgn(\kappa(s)) \kappa'(s) = |\kappa'(s)|$ for all $s < s_0$, as claimed.
\end{proof}
\subsection{Forward sweep} \label{sec:forwards}
In this subsection, the objects generated by Algorithm \ref{alg:xplus} during the forward sweep
$$x_F:= \mathtt{Sweep}(\kappa,A,C,v_0),$$
are examined under Assumptions \ref{ass:curvature} and \ref{ass:finiteA}.
Lines \ref{ln:initWhile} -- \ref{ln:endWhile} of Algorithm \ref{alg:xplus} yield the chain of inequalities
\begin{equation}
	\forall k \in [0:N],\ 0 \leq c_k \leq a_k \leq c_{k+1},\label{eq:chain}
\end{equation}
with Assumption \ref{ass:finiteA} guaranteeing the existence of
\begin{equation}  N := \max \{ k \mid c_k < \infty\}, \label{eq:N} \end{equation}
the index of the largest finite $c_k$. By this definition, $c_k,a_k \in [0,s_0)$ for each $k \leq N$, $c_{N+1} = \infty$, and $a_N \in [0,s_0) \cup \{\infty\}$.
To facilitate the subsequent analysis, define \begin{equation} \eta:[0,L] \to [0:N],\ \eta(s):= \max \{ k \in [0:N] \mid c_k \leq s\}, \label{eq:eta} \end{equation}
which returns the unique index $\eta(s)=k$ such that $s \in [c_k,c_{k+1})$.
Line \ref{ln:x+} can then be written as
\begin{align} x_F(s)  &= \begin{cases}
		C |\kappa(s)|^{-1},& \text{if } s \in \bigcup_{k=0}^N [c_k,a_k) \\
		x_{\eta(s)}(s),& \text{if } s \in \bigcup_{k=0}^N [a_k,c_{k+1})
	\end{cases} \label{eq:x+_expanded} \\
&=  \begin{cases}
	C |\kappa(s)|^{-1},& \text{if } s \in \bigcup_{k=0}^N [c_k,a_k) \\
	x_0(s),& \text{if } s \in [a_0,c_1) \\
	x_1(s),& \text{if } s \in [a_1,c_2) \\
	\vdots & \vdots \\
	x_N(s),& \text{if } s \in [a_N,c_{N+1})
\end{cases}. \label{eq:x+informal} \end{align}
\begin{remark}[Non-negativity] \label{rem:strictPos}
For all $s$, $C|\kappa(s)|^{-1}  > 0$. Moreover, since $A>0$, every $x_k(s) \geq 0$ for all $s \geq a_k$,
by Lines \ref{ln:x0} and \ref{ln:x+k}. This makes it clear from \eqref{eq:x+informal} that $x_F$ is non-negative, taking a possibly infinite value at $s_0$ when $\kappa(s_0) = 0$.
\end{remark}
First consider the output of Algorithm \ref{alg:xplus} when $s_0 = 0$, which is the simplest case.
\begin{lemma}\label{lem:t0Left}
Suppose Assumption \ref{ass:curvature} holds and $s_0 = 0$. If $v_0^2 < C|\kappa(0)|^{-1}$, then $$ \forall s \in [0,L],\ x_F(s) = v_0^2 + 2As.$$ Otherwise, $$\forall s \in [0,L],\ x_F(s) = C|\kappa(s)|^{-1}.$$
\end{lemma}
\begin{proof}
If $v_0^2 < C|\kappa(0)|^{-1}$, then $a_0 = 0$ by Line \ref{ln:a0}, and $c_1 = \infty$ by Line \ref{ln:c1}, because $[0,s_0) = \emptyset$.
Otherwise, $c_0 = 0$, and $a_0 = \infty$ by Line \ref{ln:ak}. In both cases, Assumption \ref{ass:finiteA} is satisfied automatically, and the result follows from \eqref{eq:x+_expanded}.
\end{proof}
Now consider the case $s_0 > 0$. First, a technical result.
\begin{lemma} \label{lem:finite_a'k}
	Suppose Assumption \ref{ass:curvature} holds and $s_0 > 0$. If $ c_k < \infty$, then $a_k < s_0$ or $\kappa(s_0) \neq 0$.
\end{lemma}
\begin{proof}
	Suppose that $c_k < \infty$ and $\kappa(s_0) = 0$. If $k = 0$, then $0 = c_0 < s_0$ by hypothesis; otherwise $k > 0$, and $c_k < s_0$ by Lines \ref{ln:c1} and \ref{ln:ck}. Either way, $c_k < s_0$. Recalling $\xi:[0,s_0) \to \R$ defined in Lemma \ref{lem:vdot}, $\limsup_{s \to s_0} \xi'(s) = \limsup_{s \to s_0}\frac{C |\kappa'(s)|}{\kappa(s)^2} =\infty$ by Proposition \ref{prop:deriv}, and thus, $a_k < s_0$ by Lines \ref{ln:a0} and \ref{ln:ak}.
\end{proof}
 Recall that $x_F$ enters the curvature constraint boundary at each $c_k$ and leaves it at $a_k$. In view of Lemma \ref{lem:vdot}, the tangential acceleration does not exceed $A$ while travelling along this boundary.
\begin{lemma}\label{lem:derivBounds}
	Suppose Assumption \ref{ass:curvature} holds and $s_0 > 0$. Then $\frac{C |\kappa'(s)|}{\kappa(s)^2} \leq 2A$ for all $s \in [c_k,a_k) \cap [0,L)$. Moreover, if $c_k < a_k < s_0$, then $\frac{C |\kappa'(a_k)|}{\kappa(a_k)^2} = 2A$.
\end{lemma}
\begin{proof}
	Consider the following cases. \begin{itemize}
		\item If $c_k = a_k$, then $[c_k,a_k) = \emptyset$, and the result holds vacuously.
		\item If $c_k < a_k < s_0$, then the result follows from Proposition \ref{prop:inf} with $f(s) := \frac{C |\kappa'(s)|}{\kappa(s)^2} - 2A$, which derives its continuity from that of $\kappa'$.
		\item Suppose $c_k < a_k = \infty$. Then Line \ref{ln:ak} gives
		\begin{equation}
			\forall s \in [c_k,s_0),\ \frac{C |\kappa'(s)|}{\kappa(s)^2} \leq 2A.
		\end{equation}
	Moreover, since $s_0 > 0$, it follows from Lines \ref{ln:initWhile} and \ref{ln:ck} that $c_k < s_0 \in [0,L]$. Further, since $a_k = \infty$ here, $\kappa(s_0) \neq 0$ by Lemma \ref{lem:finite_a'k}, and thus $s_0 = L$ by Corollary \ref{cor:kappaNonzero}, which establishes the result.
	\end{itemize}
This exhausts all possibilities for the values of $c_k$ and $a_k$.
\end{proof}
The normal acceleration limit $C$ is also not exceeded while travelling at maximum tangential acceleration between $a_k$ and $c_{k+1}$.
\begin{lemma} \label{lem:x+k_Rounds}
	Suppose Assumption \ref{ass:curvature} holds and $s_0 > 0$. Then $x_k(s) \leq C|\kappa(s)|^{-1}$ for all $s \in  [a_k,c_{k+1}) \cap [0,s_0]$. Moreover, if $a_k < c_{k+1} < s_0$, then $ x_k(c_{k+1}) = C|\kappa(c_{k+1})|^{-1}$.
\end{lemma}
\begin{proof}
	Consider the following cases. \begin{itemize}
		\item If $a_k = c_{k+1}$, then $[a_k,c_{k+1}) = \emptyset$, and the result holds vacuously.
		\item Suppose $a_k < c_{k+1} < s_0$. Lines \ref{ln:c1} and \ref{ln:ck} can both be written as
		$$  c_{k+1} = \inf \left \{ s \in [a_k,s_0) \mid x_k(s) -C|\kappa(s)|^{-1} > 0 \right\}.$$
		Proposition \ref{prop:inf} then implies that $x_k(s) -C|\kappa(s)|^{-1} \leq 0$ for all $s \in [a_k,c_{k+1}]$, and that $x_k(c_{k+1}) -C|\kappa(c_{k+1})|^{-1} = 0$, as claimed.
		\item If $a_k < c_{k+1} = \infty$, then $a_k < s_0$ by Lines \ref{ln:a0} and \ref{ln:ak}, whereby Lines \ref{ln:c1} and \ref{ln:ck} give
		\begin{equation} \forall s \in [a_k,s_0),\ x_k(s) \leq C|\kappa(s)|^{-1}.\label{eq:lessThanCurvature} \end{equation}
		Since $ [a_k,s_0) = [a_k,c_{k+1}) \cap [0,s_0)$, it only remains to check the inequality at $s_0$. If $\kappa(s_0) = 0$, then clearly $x_F(s_0) \leq C|\kappa(s_0)|^{-1} = \infty$. Otherwise, \eqref{eq:lessThanCurvature} implies $$ x_k(s_0) = \lim_{s \uparrow s_0}x_k(s) \leq   \lim_{s \uparrow s_0}C|\kappa(s)|^{-1} = C|\kappa(s_0)|^{-1},$$
		because $\kappa$ and $x_k$ are continuous.
	\end{itemize}
Lines \ref{ln:a0}, \ref{ln:c1}, \ref{ln:ak} and \ref{ln:ck} permit no other possibilities for the values of $a_k$ and $c_{k+1}$.
\end{proof}
It follows that $x_F$ satisfies the curvature constraint throughout the portion of the path with strictly decreasing unsigned curvature.
\begin{corollary} \label{cor:x+feasibility}
	Under Assumptions \ref{ass:curvature}--\ref{ass:finiteA}, $x_F:[0,L] \to [0,\infty]$ satisfies
	\begin{equation}
		\forall s \in [0,s_0],\ x_F(s) \leq C|\kappa(s)|^{-1}. \label{eq:x+_Rounds}
	\end{equation}
\end{corollary}
\begin{proof}
If $s_0 = 0$, then $[0,s_0] = \{0\}$, and the result follows from Lemma \ref{lem:t0Left}. Suppose $s_0 > 0$, and choose any $s \in [0,s_0]$.  Let $k := \eta(s)$, where $\eta$ is defined in \eqref{eq:eta}.  Then $s \in [c_k,c_{k+1}) = [c_k,a_k) \cup [a_k,c_{k+1})$.
	If $s \in  [c_k,a_k)$, then $x_F(s) = C|\kappa(s)|^{-1}$ by Line \ref{ln:x+}. Otherwise, $s \in [a_k,c_{k+1})$, and $x_F(s) \leq C|\kappa(s)|^{-1} $ by Lemma \ref{lem:x+k_Rounds}.
\end{proof}
The boundary condition \eqref{eq:initBC} is also satisfied by $x_F$.
\begin{lemma} \label{lem:BC}
	Under Assumptions \ref{ass:curvature}--\ref{ass:finiteA}, $x_F(0) \leq v_0^2$.
\end{lemma}
\begin{proof}
If $C|\kappa(0)|^{-1} \leq v_0^2$, then the result follows from Corollary \ref{cor:x+feasibility}. Suppose, instead, that $v_0^2 < C|\kappa(0)|^{-1} $. Then $a_0 = 0$ by Line \ref{ln:a0}. If $c_1 = 0$, then Proposition \ref{prop:inf} implies $C|\kappa(0)|^{-1} = C|\kappa(c_1)|^{-1} \leq x_0(c_1) = v_0^2$, which is a contradiction. Thus, $c_1 > 0$, $0 \in [a_0,c_1)$, and in view of \eqref{eq:x+informal}, $x_F(0) = x_0(0) = v_0^2 $ by Line \ref{ln:x0}.
\end{proof}
The next result establishes, among other things, that under conditions on $s_0$ and $v_0$, the forward speed-squared profile $x_F$ satisfies the tangential acceleration constraint wherever its derivative exists.
\begin{lemma} \label{lem:continuity}
Suppose Assumptions \ref{ass:curvature} and \ref{ass:finiteA} hold. If $s_0 > 0$ or $v_0^2 < C|\kappa(0)|^{-1}$, then $x_F$ is real-valued,  Lipschitz, strictly increasing, and
\begin{equation}
	\forall s \in (0,L) \setminus \bigcup_{k=0}^N \{c_k,a_k\},\ 0 \leq x'_F(s) \leq 2A. \label{eq:derivBounds}
\end{equation}
\end{lemma}
\begin{proof}
If $v_0^2 < C|\kappa(0)|^{-1}$ and $s_0 = 0$, then the result follows immediately from the first part of Lemma \ref{lem:t0Left}, whereby $x_F'(s) = 2A > 0$ for all $s \in [0,L]$.

Suppose, now, that $s_0 > 0$. From Lines \ref{ln:x0} and \ref{ln:x+k} of Algorithm \ref{alg:xplus}, $x_k$ is only defined if $a_k < \infty$, by which $a_k < s_0$. Since $\kappa(s) \neq 0$ for any $s \neq s_0$, every $x_k$ is real-valued. If $\kappa(s_0) \neq 0$, then $C|\kappa(s)|^{-1} < \infty$ for all $s \in [0,L]$, and so it follows from Line \ref{ln:x+} that $x_F(s) < \infty$ for all $s \in [0,L]$.
Suppose, instead, that $\kappa(s_0) = 0$, and choose any $s \in [0,L]$. Then $s \in [c_k,c_{k+1})$, where $k:= \eta(s)$. If $s \in [c_k,a_k)$, then $c_k \leq s<a_k < s_0$ by Lemma \ref{lem:finite_a'k},  and  $x_F(s) = C|\kappa(s)|^{-1} < \infty$ by Line \ref{ln:x+}. Otherwise, $s \in [a_k,c_{k+1})$, in which case $x_F(s) = x_k(s) \in \R$. This establishes that $x_F$ is real valued. The rest of the proof establishes the other properties claimed.

If $s \in (0,L) \setminus \bigcup_{k=0}^N \{c_k,a_k\} $, then either $s \in (c_k,a_k)$ or $s \in (a_k,c_{k+1})$, where $k = \eta(s)$. The latter implies that $x'_F(s) = x'_k(s) = 2A > 0 $ by \eqref{eq:x+informal}. In case of the former, $s<a_k < s_0$ by Lemma \ref{lem:finite_a'k}, and therefore $x'_F(s) =  \frac{C|\kappa'(s)|}{\kappa(s)^2} \in [0,2A]$ by Lemmas \ref{lem:vdot} and \ref{lem:derivBounds}. This establishes \eqref{eq:derivBounds}.

It is clear $x_F$ is piecewise continuous, because $\kappa$ is continuous and so is every $x_k$. All that remains is to test $x_F$ for continuity at $c_1,...,c_N$, and $a_0,...,a_{N-1}$, and finally at $a_N$ if finite. From \eqref{eq:x+_expanded}, $x_F$ is right-continuous at every finite $c_k$ and $a_k$.
If $a_k < c_{k+1} < \infty $, then by Lemma \ref{lem:x+k_Rounds},
$$ \lim_{s \uparrow c_{k+1}} x_F(s) = x_k(c_{k+1}) = C|\kappa(c_{k+1})|^{-1} = x_F(c_{k+1}),$$
which establishes left continuity at $c_1,c_2,...c_N$.
If $c_k < a_k$, then
$$ \lim_{s \uparrow a_k} x_F(s) =  C|\kappa(a_k)|^{-1} = x_k(a_k) = x_F(a_k),$$
which establishes left-continuity at every finite $a_k$. Thus $x_F$ is continuous.

To demonstrate that $x_F$ is Lipschitz, consider the right-derivative $x^+_F(s) $, which is equal to $x'_F(s)$ wherever the latter exists. By \eqref{eq:derivBounds}, $x^+_F(s) \in [0,2A]$ for all $ s \in (0,L) \setminus \bigcup_{k=0}^N \{c_k,a_k\}$. The existence of $x^+_F(s)$ at $s \in \bigcup_{k=0}^N \{c_k,a_k\} \cap [0,L)$ is now verified. If $a_k < c_{k+1}$, \eqref{eq:x+informal} implies
$ x^+_F(a_k) = x'_k(a_k) = 2A. $
Similarly, if $c_k < a_k$, then $c_k < s_0$, and with $\xi:[0,s_0) \to \R$ defined as in Lemma \ref{lem:vdot},
$$ x^+_F(c_k) = \xi'(c_k) = \frac{C|\kappa'(c_k)|}{\kappa(c_k)^2} \in [0,2A],$$
by Lemma \ref{lem:derivBounds}.
This establishes that $x^+_F(s)$ exists and is bounded for all $s \in [0,L)$, which implies $x_F$ is Lipschitz  by\cite[Proposition 2.2.3]{cobzasLipschitzFunctions2019}.

Finally, strict monotonicity is confirmed. Every $x_k$ is strictly increasing because $A>0$, and $|\kappa(s)|$ is strictly decreasing over $[0,s_0]$. If $s_0 = L$, then clearly
\begin{equation} [0,L] \cap \bigcup_{k=0}^N[c_k,a_k) \subset [0,s_0] \label{eq:domainInclusion}.\end{equation}
Otherwise $s_0 \in (0,L)$, and $\kappa(s_0) = 0$ by Lemma \ref{lem:kappa=0}. Since $c_N < \infty$ by definition in \eqref{eq:N}, Lemma \ref{lem:finite_a'k} then implies $a_N < s_0 < L$, and therefore \eqref{eq:domainInclusion} still holds. Thus, $C|\kappa(s)|^{-1}$ is strictly increasing over $[0,L] \cap \bigcup_{k=0}^N[c_k,a_k) $. Since the continuity of $x_F$ has been established, and its piecewise definition consists of only strictly increasing functions, it follows that $x_F$ is strictly increasing.
\end{proof}

\subsection{Reverse sweep} \label{sec:backwards}
Recall $\gamma:[0,L] \to [0,L],\ \gamma(s) = L - s$. If $\kappa$ satisfies Assumption \ref{ass:curvature}, then so does $\kappa \circ \gamma$, where $\gamma$ reverses direction along the path.
The profile \begin{equation} x_R:= \mathtt{Sweep}(\kappa \circ \gamma,B,C,v_L)\circ \gamma \label{eq:x-alg} \end{equation}
is obtained by applying the smooth co-ordinate transformation $(s \mapsto L - s)$ to both input and output of Algorithm \ref{alg:xplus}. The relevant properties of $x_R$ follow directly from those of $x_F$ established previously. To avoid confusion, internal variables of Algorithm \ref{alg:xplus} generated by the reverse sweep~\eqref{eq:x-alg} are adorned with a tilde.
For example,
$$ \tilde{s}_0:= \argmin_{s \in [0,L] }|\kappa \circ \gamma(s)| = \gamma(s_0) = L - s_0,$$
where $s_0 := \argmin_{s \in [0,L] }|\kappa(s)|$, as per \eqref{eq:t0}. The next four results are corollaries of Lemma \ref{lem:t0Left}, Corollary \ref{cor:x+feasibility}, Lemma \ref{lem:BC}, and Lemma \ref{lem:continuity}, respectively.
\begin{corollary} \label{cor:t0=L}
	Suppose Assumption \ref{ass:curvature} holds, and $s_0 = L$. If $v_L^2 < C|\kappa(L)|^{-1}$, then
	$$ \forall s \in [0,L],\ x_R(s) = v_L^2 + 2B(L-s).$$
	 Otherwise, $$ \forall s \in [0,L],\ x_R(s) = C|\kappa(s)|^{-1}.$$
\end{corollary}
\begin{corollary} \label{cor:x-feasibility}
	Under Assumptions \ref{ass:curvature} and \ref{ass:finiteB}, $x_R:[0,L] \to [0,\infty]$ satisfies
	\begin{equation*}
		\forall s \in [s_0,L],\ x_R(s) \leq C|\kappa(s)|^{-1}.
	\end{equation*}
\end{corollary}
\begin{corollary} \label{cor:x-BC}
	Under Assumptions \ref{ass:curvature} and \ref{ass:finiteB}, $x_R(L) \leq v_L^2$.
\end{corollary}
\begin{corollary} \label{cor:x-continuity}
	Suppose Assumptions \ref{ass:curvature} and \ref{ass:finiteB} hold. If $s_0 < L$ or $v_L^2 < C|\kappa(L)|^{-1}$, then $x_R$ is real-valued, Lipschitz, strictly decreasing, and
	\begin{align*}
		\forall s \in (0,L) \setminus \bigcup_{k=0}^N \{L-\tilde{a}_k,L-\tilde{c}_k\},\ -2B \leq x'_R(s) \leq 0.
	\end{align*}
\begin{proof}
The map $\gamma$ is Lipschitz, and the composition of Lipschitz functions remains Lipschitz~\cite[Proposition 2.3.1]{cobzasLipschitzFunctions2019}.
\end{proof}
\end{corollary}
\begin{remark}[Relationship to \cite{consoliniSolutionMinimumtimeSpeed2020}]
	\label{rem:link}
In view of the constant limits and strictly-monotone path curvature $\kappa$, it can be shown that the profiles $x_F$ and $x_R$ solve the respective discontinuous ODEs \cite[(11-12)]{consoliniSolutionMinimumtimeSpeed2020}, with
\begin{align*}
	& \alpha^+(s) := 2A, \\ 
	& \alpha^-(s) := -2B,\\
	& \mu(s) := C|\kappa(s)|^{-1}.
\end{align*}
It is conceivable that the monotonicity assumption could be relaxed without sacrificing this property; a possibility that warrants further investigation.
\end{remark}
\subsection{Feasibility and global optimality}
Recall the definition of $x^\star$ in Theorem \ref{thm:soln}:
\begin{equation} x^\star:[0,L] \to [0,\infty),\ x^\star(s) := \min \{ x_F(s), x_R(s), V^2 \}. \label{eq:xstar} \end{equation}
The feasibility of $x^\star$ is established first, followed by global optimality.
\begin{lemma}[Feasibility] \label{lem:feasible}
	Under Assumptions \ref{ass:curvature} -- \ref{ass:finiteB}, $x^\star$ in Theorem \ref{thm:soln} is Lipschitz and feasible.
\end{lemma}
\begin{proof}
By definition, $x^\star(s) \leq V^2$ for all $s \in [0,L]$. Moreover, $x^\star:[0,L] \to [0,\infty)$, because $x_F,x_R:[0,L] \to [0,\infty]$ and $0<V<\infty$. To verify \eqref{eq:initBC}--\eqref{eq:finalBC}, $x^\star(0) \leq x_F(0) \leq v_0^2$ by Lemma \ref{lem:BC}, and $x^\star(L) \leq x_R(L) \leq v_L^2$ by Corollary \ref{cor:x-BC}.
Now choose any $s \in [0,L]$. If $s \leq s_0$, then $x^\star(s) \leq x_F(s) \leq C|\kappa(s)|^{-1}$ by Corollary \ref{cor:x+feasibility}. Otherwise, $s > s_0$, and $x^\star(s) \leq x_R(s) \leq C|\kappa(s)|^{-1}$ by Corollary \ref{cor:x-feasibility}. Thus, \eqref{eq:stateConstraints} is verified.

To verify \eqref{eq:ocpDynamics} and absolute continuity of $x^\star$, consider the following cases.
\begin{itemize}
	\item Suppose $s_0 = 0$ and $v_0 \geq C|\kappa(0)|^{-1}$. Then $x_F(s) = C|\kappa(s)|^{-1}$ for all $s \in [0,L]$, by Lemma \ref{lem:t0Left}. Since $x_R(s) \leq C|\kappa(s)|^{-1} = x_F(s)$ for all $s \in [0,L]$ by Corollary \ref{cor:x-feasibility}, \begin{equation} \forall s \in [0,L],\ x^\star(s) = \min\{ x_R(s),V^2\}, \label{eq:xstar-} \end{equation}
	where $x_R$ is Lipschitz and strictly decreasing by Corollary \ref{cor:x-continuity}, with $x'_R(s) \in [-2B,0]$ almost everywhere.

\item Suppose $s_0 = L$ and $v_L \geq C|\kappa(L)|^{-1}$. Then $x_R(s) = C|\kappa(s)|^{-1}$ for all $s \in [0,L]$ by Corollary \ref{cor:t0=L}. Since $x_F(s) \leq C|\kappa(s)|^{-1} = x_R(s)$ for all $s \in [0,L]$ by Corollary \ref{cor:x+feasibility}, \begin{equation} \forall s \in [0,L],\ x^\star(s) = \min\{ x_F(s),V^2\}, \label{eq:xstar+} \end{equation}
where $x_F$ is Lipschitz and strictly increasing by Lemma \ref{lem:continuity}, with $x'_F(s) \in [0,2A]$ almost everywhere.

\item Otherwise, $s_0 \in (0,L)$ or $v_0 < C|\kappa(0)|^{-1}$ or $v_L < C|\kappa(L)|^{-1}$. Then $x_F$ is strictly increasing with $x'_F(s) \in [0,2A]$ almost everywhere by Lemma \ref{lem:continuity}, and $x_R$ strictly decreasing with $x'_R(s) \in [-2B,0]$ almost everywhere by Corollary \ref{cor:x-continuity}. Both $x_F$ and $x_R$ are Lipschitz. Equation \eqref{eq:xstar} can be written as
$ x^\star(s) = \min\left\{ x_R(s), g(s) \right\} $, where $g(s):= \min \{x_F(s), V^2 \}$ is non-decreasing.
\end{itemize}
Applying Proposition \ref{prop:derivAC} to each case, $x^\star$ is Lipschitz and satisfies \eqref{eq:ocpDynamics} almost everywhere. All real Lipschitz functions over a compact domain are absolutely continuous~\cite[Proposition 3.3.3]{cobzasLipschitzFunctions2019}.
\end{proof}

\begin{lemma} \label{lem:lessthan}
Suppose Assumptions \ref{ass:curvature} and \ref{ass:finiteA} hold, and let $x:[0,L] \to \R$ be absolutely continuous. If
\begin{align}  &x(0) \leq v_0^2, \label{eq:v0} \\
\forall s \in [0,L],\ &x(s) \leq C|\kappa(s)|^{-1}, \label{eq:cBound}
\end{align}
 and $x'(s) \leq 2A$ for almost every $s \in [0,L]$, then
$x(s) \leq x_F(s)$ for all $s \in [0,L]$.
\end{lemma}
\begin{proof}
	Choose any $s \in [0,L]$. Then $s \in [c_k,c_{k+1})$, where $k:= \eta(s)$. If $s \in [c_k,a_k)$, then \eqref{eq:cBound} and \eqref{eq:x+_expanded} imply $x(s) \leq C|\kappa(s)|^{-1} = x_F(s) $. Suppose, instead, that $s \in [a_k,c_{k+1})$. If $k = 0$ and $v_0^2 < C|\kappa(0)|^{-1}$, then $a_k = 0$ by Line \ref{ln:a0}, and \eqref{eq:v0} implies
	$$ x(a_0) = x(0) \leq v_0^2 = x_0(0),$$
	by Line \ref{ln:x0}. Otherwise, \eqref{eq:cBound} implies $$  x(a_k) \leq C|\kappa(a_k)|^{-1} =  x_k(a_k) ,$$
	by Line \ref{ln:x+k}. Either way, $x(a_k) \leq x_k(a_k)$. Since $x$ is absolutely continuous with  $x'(s) \leq 2A$ almost everywhere,
	\begin{align*}
		x(s) &= x(a_k) + \int_{a_k}^s x'(\tau) \d \tau  \\
		& \leq x_k(a_k) +  2\int_{a_k}^s A \d \tau \\
		& =  x_k(a_k)+ 2A(s - a_k) \\
		& = x_k(s) = x_F(s),
	\end{align*}
by \eqref{eq:x+informal}, recalling that $ s \in [a_k, c_{k+1})$.
\end{proof}
\begin{corollary} \label{cor:lessThan}
	Suppose Assumptions \ref{ass:curvature} and \ref{ass:finiteB} hold, and let $x:[0,L] \to \R$ be absolutely continuous. If \begin{align*}
		&x(L) \leq v_L^2, \\
		\forall s \in [0,L],\ &x(s) \leq C|\kappa(s)|^{-1},  \end{align*} and $x'(s) \geq -2B$ for almost every $s \in [0,L]$, then
	$x(s) \leq x_R(s)$ for all $s \in [0,L]$.
\end{corollary}
\begin{proof}
Recall $\gamma(s):= L-s$. If $x$ satisfies the hypotheses, then $x \circ \gamma$ satisfies $(x \circ \gamma)'(s) \leq 2B$ almost everywhere, $x \circ \gamma(0) \leq v_L^2$, and \eqref{eq:cBound}. Thus $x \circ \gamma(s) \leq \tilde{x}_F(s)$ for all $s$ by Lemma \ref{lem:lessthan}, where $\tilde{x}_F(s) := \mathtt{Sweep}(\kappa \circ \gamma, B, C,v_L)$. Since $x_R = \tilde{x}_F \circ \gamma$, the result follows.
\end{proof}
\begin{corollary} \label{cor:lessThanStar}
Under Assumptions \ref{ass:curvature}--\ref{ass:finiteB}, for any feasible $x$, $x(s) \leq x^\star(s)$ for all $s \in [0,L]$.
\end{corollary}
\begin{proof}
This follows directly from Lemma \ref{lem:lessthan}, Corollary \ref{cor:lessThan}, Definition \ref{def:feasible} and \eqref{eq:xstar}.
\end{proof}
\begin{corollary}[Global optimality]
Under Assumptions \ref{ass:curvature}--\ref{ass:finiteB}, $J(x) \geq J(x^\star)$ for any feasible $x$.
\end{corollary}
\begin{proof}
If $x:[0,L] \to \R$ is feasible, Corollary \ref{cor:lessThanStar} implies $x(s) \leq x^\star(s)$ for all $s$, and the result follows because the running cost $\frac{1}{\sqrt{x(s)}} $ in \eqref{eq:cost} is a strictly decreasing function of $x(s)$.
\end{proof}
Theorem \ref{thm:soln} has been proved. Now consider tightening boundary conditions \eqref{eq:initBC}, \eqref{eq:finalBC}, or both, to equality in Problem \ref{prob:OCP}. If a solution exists, then such is $x^\star$.
\begin{corollary}[Equality-constrained boundary values] \label{cor:BCs}
	Under the hypotheses of Theorem \ref{thm:soln}: \begin{enumerate}[i)]
		\item If there exists a feasible speed-squared profile $x$ with $x(0) = v_0^2$, then $x^\star(0) = v_0^2$.
		\item If there exists a feasible speed-squared profile $x$ with $x(L) = v_L^2$, then $x^\star(L) = v_L^2$.
	\end{enumerate}
\end{corollary}
\begin{proof}
By Corollary \ref{cor:lessThanStar}, $x(0) \leq x^\star(0) \leq v_0^2$ and $x(L) \leq x^\star(L) \leq v_L^2$.
\end{proof}

\section{Numerical example}

\begin{figure*}[p]
	\begin{subfigure}{0.5\textwidth}
		\includegraphics[width = \textwidth]{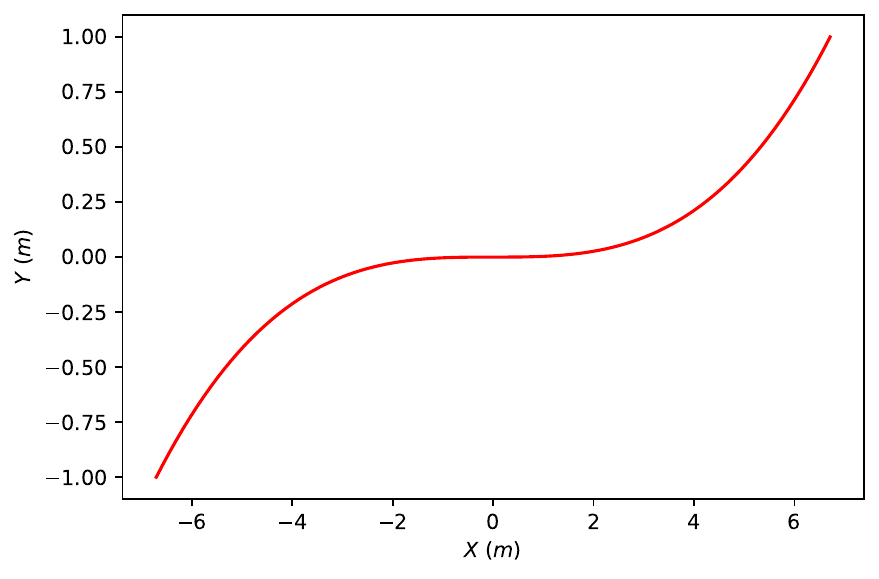}
		\caption{Planar cubic}
	\end{subfigure}
	\begin{subfigure}{0.5\textwidth}
	\includegraphics[width = \textwidth]{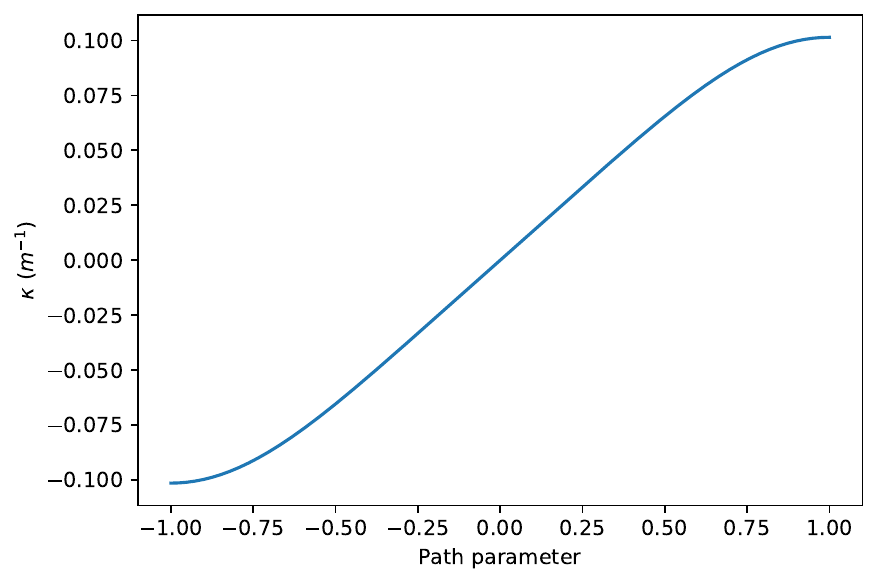}
	\caption{Signed curvature}
\end{subfigure}
	\caption{Cubic path with strictly increasing signed curvature. \label{fig:cubic}}
\end{figure*}
\begin{figure*}[p]
	\begin{subfigure}{0.5\textwidth}
		\includegraphics[width = \textwidth]{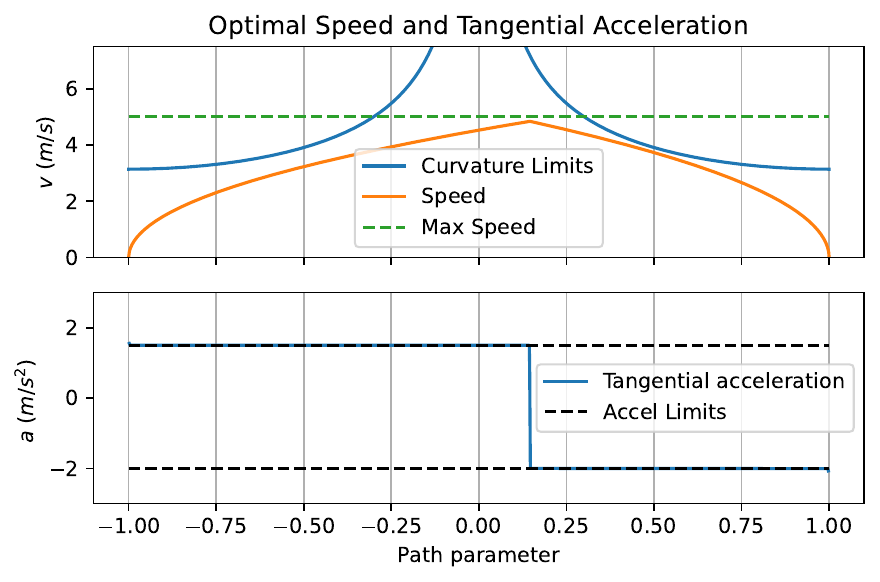}
		\caption{Boundary values: $v_0 = v_L = 0$ m/s \label{fig:zero}}
	\end{subfigure}
	\begin{subfigure}{0.5\textwidth}
	\includegraphics[width = \textwidth]{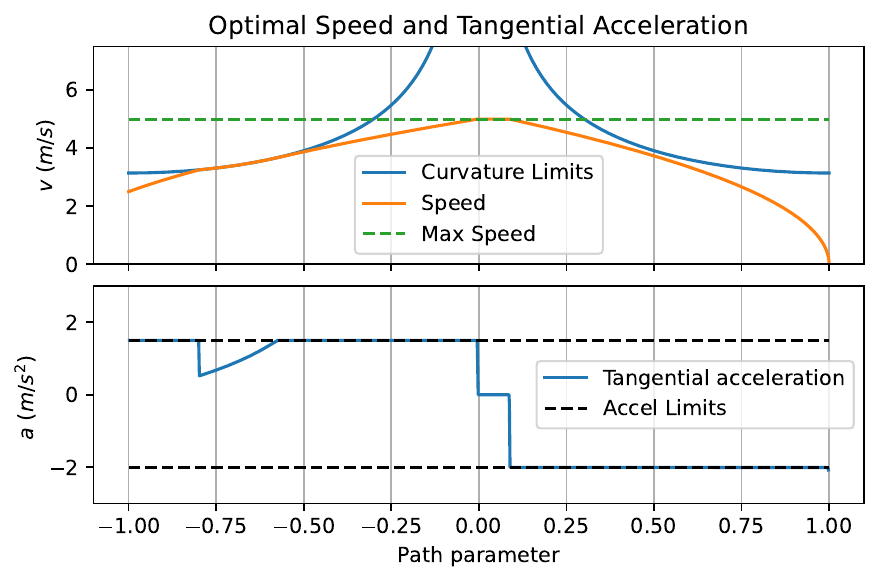}
	\caption{Boundary values: $v_0 = 2.5$ m/s, $v_L = 0$ m/s\label{fig:mixed}}
\end{subfigure}
	\begin{subfigure}{0.5\textwidth}
	\includegraphics[width = \textwidth]{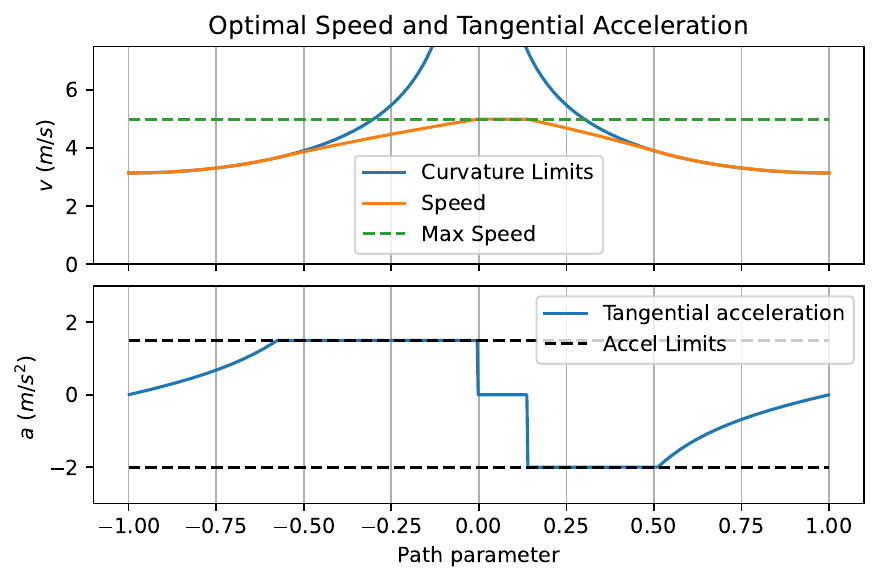}
	\caption{Free boundary values \label{fig:free}}
\end{subfigure}
	\begin{subfigure}{0.5\textwidth}
		\includegraphics[width = \textwidth]{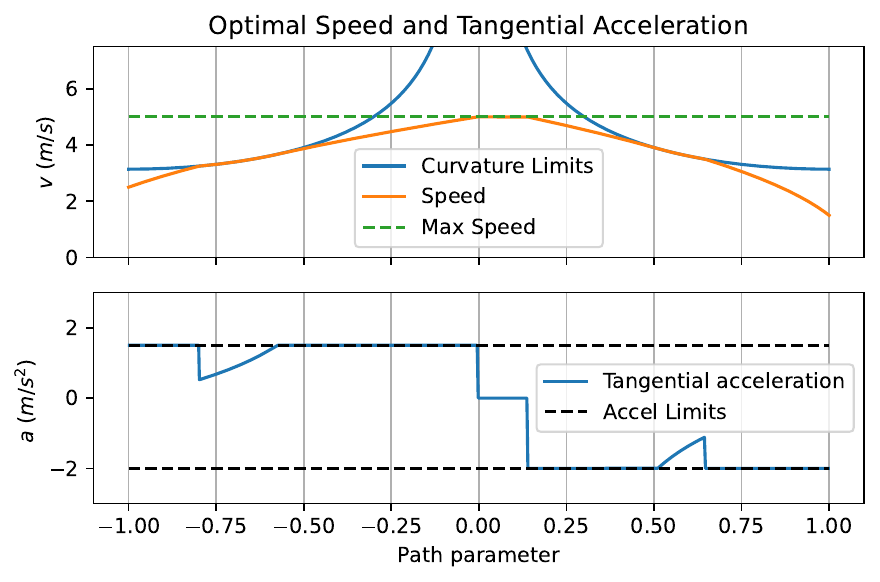}
		\caption{Boundary values: $v_0 = 2.5$ m/s, $v_L = 1.5$ m/s\label{fig:bound}}
	\end{subfigure}
	\caption{Time-optimal speed and tangential acceleration profiles (for $A = 1.5\ \mathrm{m/s^{2}},\ B = 2\ \mathrm{m/s^{2}},\ C = 1\ \mathrm{m/s^{2}},\ V = 5\ \mathrm{m/s}$). \label{fig:speed_profiles}}
\end{figure*}
Figure \ref{fig:cubic} depicts the cubic path
$$r:[-1,1] \to \R^2,\ r(\tau) = \begin{bmatrix}
	 \beta \tau \\ \tau^3
	\end{bmatrix},$$
with $\beta := 3\sqrt{5}$ chosen to ensure strictly increasing signed curvature over its domain. It is an example of a low-order clothoid approximation. Although $r$ is not parametrised by arc-length, the re-parametrisation
$$ \left( \tau \mapsto s = \int_{-1}^\tau \| r'(y) \| \d y \right) $$
can be performed numerically, before invoking Theorem \ref{thm:soln} to obtain optimal speed and tangential acceleration profiles. The profiles are plotted in Figure \ref{fig:speed_profiles}, under the stated speed and acceleration limits, and for a variety of boundary conditions. As is characteristic of time-optimal control, at every point along the path, some constraint is active. The simplest profile, obtained by setting both boundary speeds to zero, is plotted in Figure \ref{fig:zero}. Only a single switch is required, from maximum tangential acceleration to maximum deceleration. In Figure \ref{fig:mixed}, the initial speed is increased, causing the speed profile to make contact with the normal acceleration and maximum speed boundaries. The profile in Figure \ref{fig:free} is obtained by leaving both boundary speeds free (see Remark \ref{rem:BCs}). The profile with the greatest number of switches is plotted in Figure \ref{fig:bound}. It begins at maximum tangential acceleration and finishes at maximum deceleration, to achieve the boundary speeds imposed.

\section{Conclusion}
Problem \ref{prob:OCP} encodes the minimum-time speed planning task subject to limits on tangential acceleration and deceleration, normal acceleration, and speed. Boundary value constraints can be removed by setting sufficiently high boundary speeds. Theorem \ref{thm:soln} presents an exact solution for paths with smooth strictly-monotonic signed curvature, under the mild regularity conditions in Assumptions \ref{ass:finiteA} and \ref{ass:finiteB}. Similar regularity conditions appear in \cite{bobrowTimeOptimalControlRobotic1985,shinMinimumtimeControlRobotic1985,xuanphuMethodOrientingCurves1997}. The solution is constructed explicitly by invoking Algorithm \ref{alg:xplus} in two sweeps. The forward sweep progresses in the direction of increasing path parameter, to yield a speed-squared profile that switches between the maximum tangential acceleration and normal acceleration boundaries. The reverse sweep proceeds in the opposite direction to give a speed-squared profile that switches between the maximum deceleration and normal acceleration boundaries. The pointwise minimum of the two profiles and the speed-squared limit is globally time-optimal. In particular, any feasible speed-squared profile is uniformly less than or equal to this optimiser, thereby taking at least as much time to traverse. All paths with strictly-monotonic unsigned curvature have strictly-monotonic signed curvature, and can therefore be treated as special cases. Future work will address the decomposition of general paths into segments of monotone curvature, and the choice of boundary conditions to stitch segment profiles together, without losing exactness or optimality. Potential relaxation of the segment monotonicity requirements are also in view.
\section*{Acknowledgements}
Support was received from the Australian Government through Trusted Autonomous Systems, a Defence Cooperative Research Centre under the Next Generation Technologies Fund.

\bibliography{../References/FormalMethodsInControl}
\bibliographystyle{ieeetr}

\end{document}